\documentclass[journal]{IEEEtran}

\usepackage{cite}
\usepackage{graphicx,epsf,epsfig}

\usepackage{subfigure}
\usepackage{color}
\usepackage{amsmath}
\usepackage{color}
\usepackage{array}
\usepackage{fixltx2e}
\usepackage[normalem]{ulem}
\usepackage{flushend}
\usepackage{epstopdf}

\begin{document}
%
\title{Reducing error rates in straintronic multiferroic \textcolor{black}{dipole-coupled} nanomagnetic logic by pulse shaping}

\author{Kamaram Munira, Yunkun Xie, Souheil Nadri, Mark B. Forgues,  Mohammad Salehi Fashami, \\Jayasimha Atulasimha, 
Supriyo Bandyopadhyay and Avik W. Ghosh
\thanks{K. Munira is with Center for Materials for Information Technology(MINT), University of Alabama, Tuscaloosa,
AL 35401, USA. e-mail:kmunira@mint.ua.edu.}

\thanks{Y. Xie, S. Nadri, M. Forgues and A.W. Ghosh are with Charles L. Brown Department of Electrical and
 Computer Engineering, University of Virginia, Charlottesville,
VA 22903, USA.}
\thanks{M. Salehi Fashami and J. Atulasimha are with the Department of Mechanical and Nuclear Engineering, Virginia Commonwealth 
University, Richmond,
VA 23220, USA.}
\thanks{S. Bandyopadhyay is with the Department of Electrical and Computer Engineering, Virginia Commonwealth University, Richmond,
VA 23220, USA.}
\thanks{Manuscript received \today.}}

%



\maketitle

\begin{abstract}
\textcolor{black}{Dipole-coupled} nanomagnetic logic (NML), where nanomagnets with bistable 
magnetization states act as binary switches \textcolor{black}{and information is 
transferred between them via dipole-coupling and Bennett clocking}, is a potential replacement for conventional
transistor logic since magnets dissipate less energy than transistors when they switch \textcolor{black}{in response
to the clock}.
However, dipole-coupled NML is  much more error-prone than transistor logic because
thermal noise can easily disrupt magnetization dynamics. Here, we study a particularly energy-efficient 
version of dipole-coupled NML known as straintronic multiferroic logic (SML) where magnets are clocked/switched with
electrically generated mechanical strain. 
By appropriately `shaping' the voltage pulse that generates strain,
the error rate in SML can be reduced to tolerable limits. 
In this paper, we describe the error probabilities 
associated with various stress pulse shapes and discuss the trade-off between error rate 
and switching speed in SML.
\end{abstract}


%
\IEEEpeerreviewmaketitle

\section{Introduction}

\IEEEPARstart{D}{}ipole-coupled nanomagnetic logic (NML) has attracted attention as a viable paradigm for 
non-volatile logic
 because of its potential energy advantage over transistors \cite{bandy_review,Imre13012006, 
 Cowburn25022000}. In one version of \textcolor{black}{dipole-coupled} NML --
known as {\it straintronic multiferroic logic} (SML) -- magnets 
are switched  by straining a multiferroic nanomagnet with a small voltage ($\sim$10 mV), 
dissipating only an estimated 100 $kT$ of energy per magnet at room temperature (1 $kT$ = 4$\times 10^{-21}$ Joules
at room temperature) \cite{roy:063108}. 
In contrast, a modern-day transistor dissipates roughly 10,000 $kT$  to switch in isolation and about 10$^5$ $kT$ 
to switch 
in a circuit \cite{cmos}. 

\begin{figure}[ht]
\centerline{\epsfig{figure=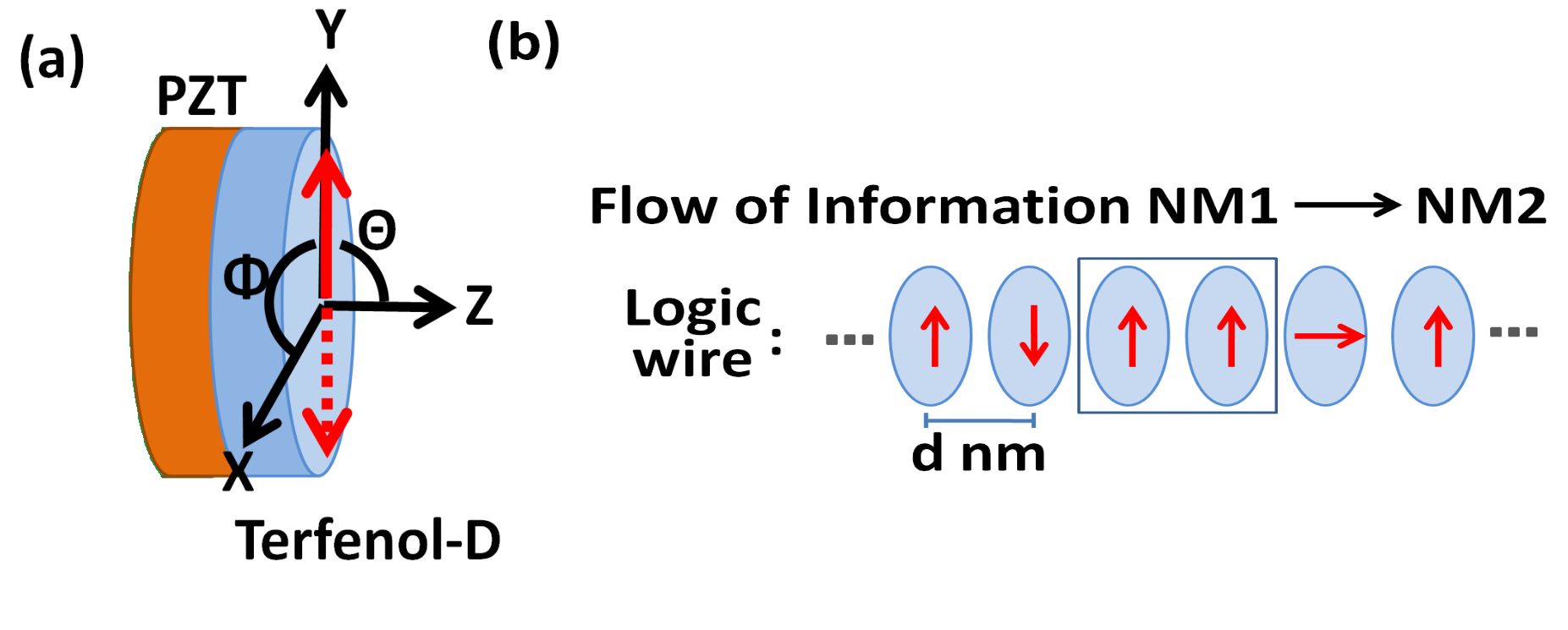,width=3.5in,height=1.2in}}
\caption{(a) A multiferroic nanomagnet (NM) (Terfenol-D/PZT). The two stable magnetization
orientations along the easy axis (major axis)  
of the elliptical Terfenol-D layer encodes bits \textquoteleft 0' ($\phi$= $\pi$/2) and \textquoteleft 1' 
($\phi$= -$\pi/2$ or 3$\pi$/2). (b) A binary wire,
spaced d nm apart.} 
\label{band_structure}
\end{figure}

While the  energy advantage of SML has been discussed extensively,  
its reliability at room temperature \textcolor{black}{has been inadequately examined}. 
Ref. \cite{spedalieri} studied room-temperature error probabilities in dipole-coupled NML logic 
clocked with magnetic fields instead of strain \textcolor{black}{and found them} to be impractically high ($>$ 1\%).
 We showed that SML with inter-magnet dipole-coupling is similarly error prone \cite{6632926}
\textcolor{black}{owing} to the out-of-plane excursion of the magnetization 
 vector during switching that \textcolor{black}{produces} a detrimental torque \textcolor{black} {which hinders correct switching} \cite{Fashami2013}. 
 Here, we will address the question of improving the reliability of \textcolor{black}{dipole-coupled SML
 by making} information transfer between two multiferroic nanomagnets (NMs) interacting 
 via dipole coupling \textcolor{black}{more robust}. We achieve this by 
 optimizing the stress profiles in time-domain, which we refer to as {\it pulse shaping}.
\textcolor{black}
 {Ref. \cite{roy:recent}
 proposed reducing error probability 
 by using a feedback circuit 
 to determine when to withdraw stress on the magnet. The feedback circuit dissipates so much energy 
 that it defeats the very purpose of SML. It is therefore an ineffective countermeasure.  We do not 
 use any such construct  and retain the energy advantage of SML.} 

 Each NM that we consider is a synthetic multiferroic stack with two layers -  a magnetostrictive (Terfenol-D) 
and a piezoelectric 
(PZT) layer (Fig. 1a). Each layer is shaped like an elliptical cylinder. In the absence of any strain, 
the two (mutually anti-parallel) orientations
 along the major axis (the so-called `easy axis') of the ellipse
($\theta = \pi/2$ and $\phi$ = $\pi/2$ or $3\pi/2$ in Fig. 1a) are stable magnetization orientations
(degenerate potential energy minima of the magnet) that encode the binary bits  
\textquoteleft 0' and  \textquoteleft 1' \cite{PhysRevB.83.224412}. They are separated in energy by
the shape anisotropy energy barrier [due to the anisotropic (elliptical) shape of the 
magnet] which prevents thermally-activated random and spontaneous switching between the two states. 
The barrier height is minimum when the magnetization vector lies in the plane of the magnet.
We will call that the in-plane shape anisotropy barrier. Strain depresses this barrier and 
causes switching of the magnetization 
from one stable state to the other via an internally generated torque \cite{roy:unpublished}.

In this paper, instead of simulating \textcolor{black}{logic bit propagation down a chain of sequentially clocked NM-s 
acting as a
a} binary wire, we will focus on the simplest dipole-coupled system of just
two multiferroic nanomagnets NM1 and NM2, and study unidirectional information transmission from NM1 acting as input
to NM2 acting as output (Fig. 1b). As long as the dipole coupling between the two magnets 
is not too strong, their magnetizations will be mutually anti-parallel in the ground state 
and hence the 
system will act as a simple inverter or NOT gate since the input and output bits are logic complements.
If we flip the input magnetization, we will expect the output to flip in response,
but this may not happen since the magnetization in NM2 may not be
able to overcome the in-plane shape anisotropy barrier to switch (because the 
dipole interaction energy is much smaller than the shape anisotropy barrier). In other words,
the system may remain stuck in a metastable state (NM1 and NM2 parallel)
instead of reaching the ground state (NM1 and NM2 anti-parallel). To break the logjam, NM2 
is mechanically stressed to depress the shape anisotropy barrier, and when the stress is 
withdrawn, it should flip to assume the anti-parallel configuration owing to dipole interaction.
This strategy is known as 
Bennett clocking \cite{bennett} which successfully transmits bit information unidirectionally
from NM1 to NM2. In a long chain of many magnets, information is propagated unidirectionally from left to right by sequentially
stressing the magnets to the right of the input magnet pairwise to rotate their magnetizations 
by large angles \cite{atulasimha:173105,noel2011}.

 Fig. 2a shows the 3-step information transmission process in the inverter. In step A, 
information has just been written in NM1 (by flipping its magnetization with an external agent). 
In step B,  NM2 is stressed (Bennett clocked) causing its magnetization to rotate. In step C, the stress on
NM2 is withdrawn whereupon its magnetization should relax to the easy axis with the final orientation {\it anti-parallel} 
to that of NM1 because of dipole coupling \cite{0957-4484-22-15-155201}. \textcolor{black}{If this fails to happen, an error is incurred}. For the sake of completeness, we should also consider the case
when the input to NM1 is such that it does not flip.
Since the Bennett clocking is performed irrespective of the input state, the stress cycle in
this case must not flip NM2; any flipping 
will result in an error.
We do not discuss this case here since it is less error-prone than the other case. We will explore the switching profiles outlined
in Fig. 2b, and study their switching reliability and delay.
\textcolor{black}{We note that the strain in the PZT responds to the applied voltage in
timescales $<$ 100 ps \cite{300668}. Since our voltage pulse widths are 1 ns or more, we can neglect effects 
associated with finite rise and fall times of the stress in response to an abrupt voltage pulse.}

Returning to the first case, successful switching of NM2 when NM1 is flipped is aided by two factors: 
(i) we need a high stress in step B to 
kick the magnetization of NM2 out of the initial orientation ($\phi$ = $\pi/2$ or $3\pi/2$) which are called 
stagnation points 
since the net 
torque on the magnetization due to stress vanishes at these precise locations, and (ii) we also need strong dipolar coupling
between the two magnets in steps B and C when the magnetization of NM2 is near the in-plane hard axis 
($\theta = \pi/2$ and $\phi$ = 0 or $\pi$) 
so that the final orientation of NM2 upon stress release is anti-parallel to that of 
NM1. The stressing profile (stress versus time) for NM2 has to be designed with 
the above two facts in 
mind to minimize the switching error.

\begin{figure}
\subfigure{\includegraphics[width=3.5in,height=1.8in]{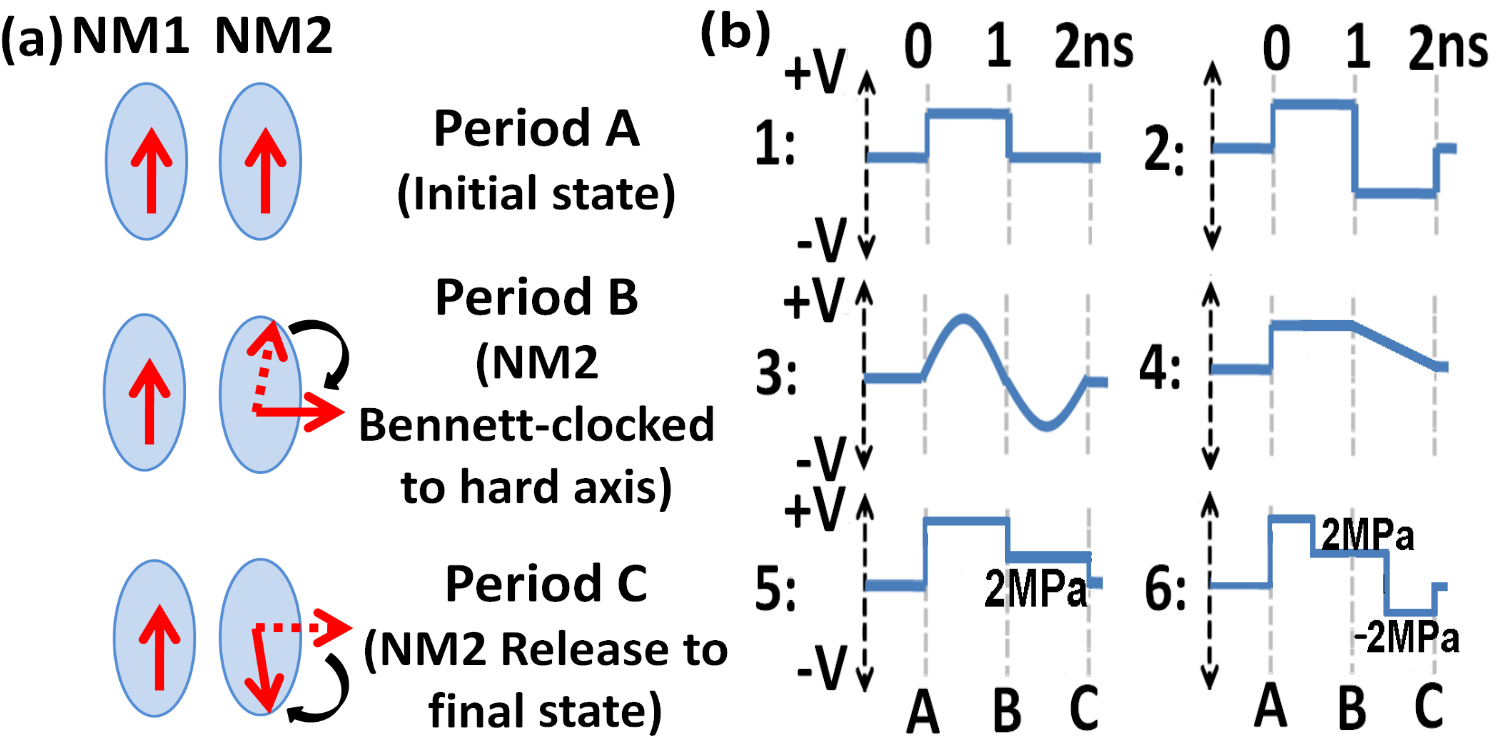}}
\subfigure{\includegraphics[width=1.72in,height=2in]{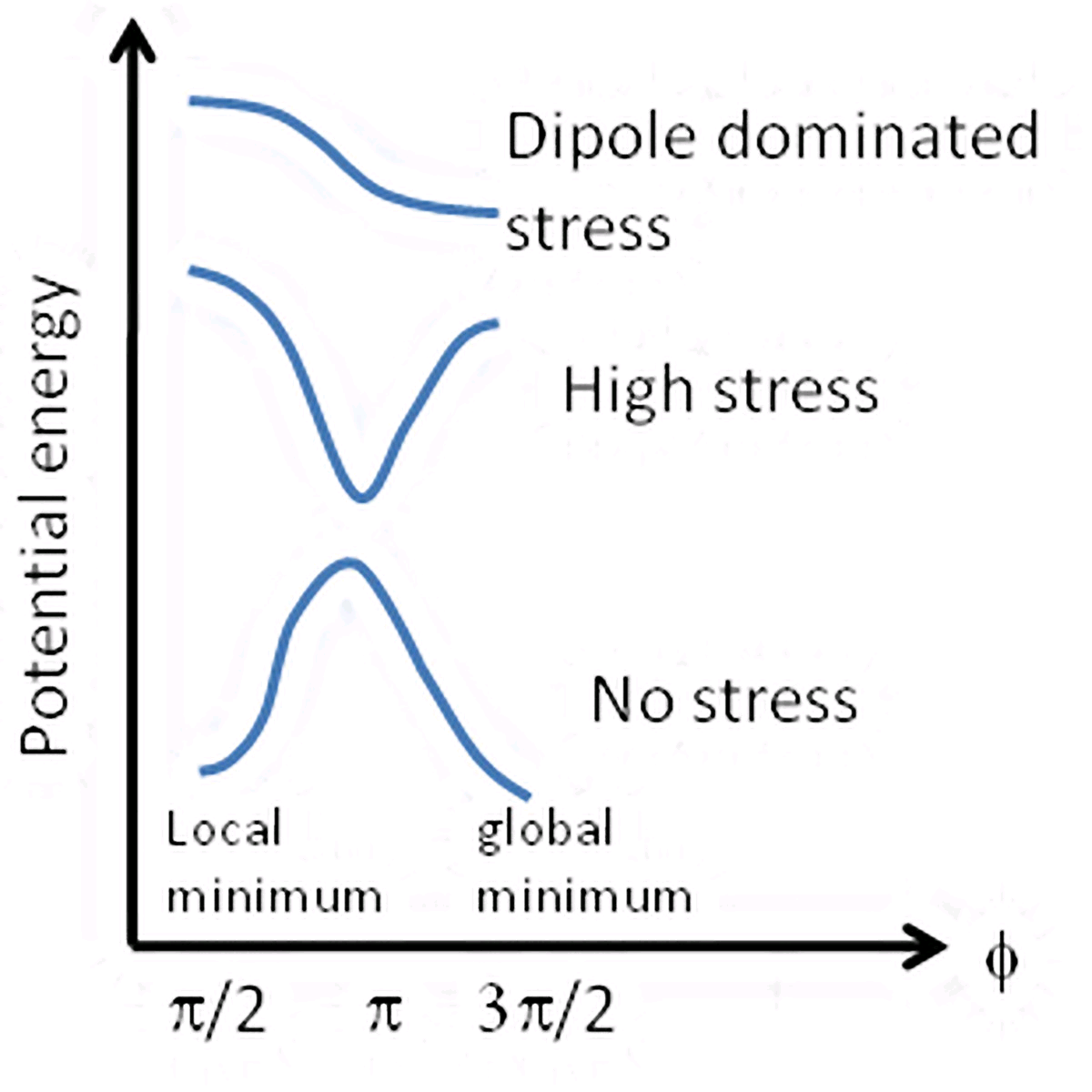}}
\subfigure{\includegraphics[width=1.72in,height=2in]{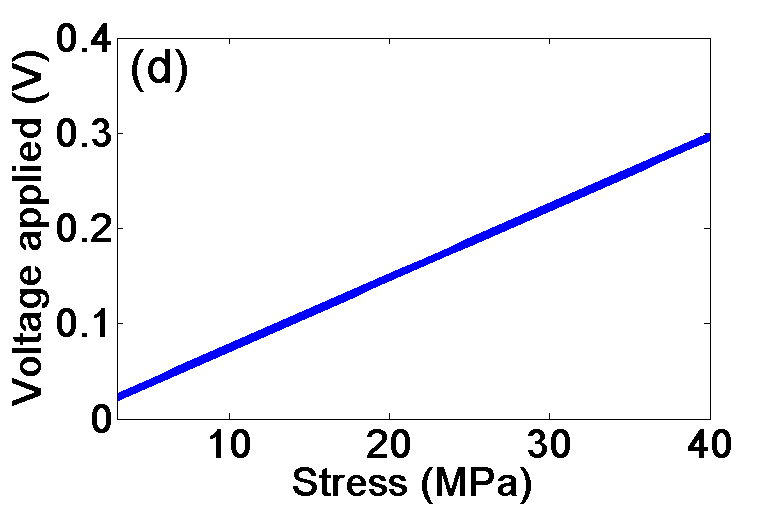}}\\\
\caption{(a) 3-step information transmission process in the two-NM inverter.  (b) Stress profiles with abrupt removal (Cases 1 and 2), 
tapered (Cases 3 and 4),  and engineered to operate mostly in the dipole dominated regime or Region II  (Cases 5 and 6). 
(c) A schematic of the potential
energy profiles ($E$ versus $\phi$) of NM2 under two different stresses. The ``up'' configuration is 
$\phi = \pi/2$ and the down configuration is $\phi = 3 \pi/2$. (d) Voltage applied to generate specified stress.} \label{band_structure}
\end{figure}

The minimum stress needed to kick the magnetization of NM2 out of a stable state along
the easy axis is the {\it critical stress}. It is roughly that value of stress where the stress anisotropy energy 
offsets the in-plane shape anisotropy energy barrier.
There is a range of stresses just above the critical stress where the  
stress anisotropy energy not only offsets the in-plane shape anisotropy energy barrier in NM2 but 
\textcolor{black}{ensures that} the only 
asymmetry in the potential energy landscape is due to the dipole coupling with 
NM1 (see upper curve of
Fig. 2c). We call this the ``dipole dominated region''.
The asymmetry will reliably switch NM2 from $\phi = \pi/2$ (``up'' in
Fig. 2a) to $\phi = 3 \pi/2$ (``down'' in Fig. 2a) because there is only one energy minimum 
and it is
located at $\phi = 3 \pi/2$. 
At higher stresses, the minimum energy location moves to
$\phi = \pi$ (see the middle curve of Fig. 2c) and the magnetization will align along the minor axis of 
the elliptical magnet. 
 When stress is finally withdrawn, the location $\phi = \pi$ will become the 
energy maximum and there will be a global minimum at $\phi = 3 \pi/2$ (ground state)
but there is also a local one at $\phi =  \pi/2$ (metastable state) as shown in the bottom curve of
Fig. 2c. Dipole interaction 
will make the global minimum lower in energy than the local minimum. Consequently,
the magnetization
 will have some preference 
for the ground state ($\phi = 3 \pi/2$) over the metastable state ($\phi =  \pi/2$),
but thermal perturbations can take the system to the (wrong) metastable state. If it ends up there, then
the in-plane
shape anisotropy energy barrier will prevent it from reaching the global minimum, thus causing 
an error.
Therefore, operating in the dipole dominated region reduces error rate but increases switching 
delay (because the stress is relatively weak), while operating at stress levels much above the dipole dominated 
region has the opposite effect. Fig. 2d shows the voltage needed to generate a 
certain amount of stress in the Terfenol-D NM, \textcolor{black}{assuming complete strain transfer 
from the PZT to the Terfenol-D}. The voltage was estimated by taking the PZT layer thickness to be 40 nm, Young's 
modulus of Terfenol-D to be 30 GPa, and the d$_{31}$ coefficient of \textcolor{black}{PZT}
 to be 1.8x10$^{-10}$ m/V. 
\textcolor{black}{All stress quantities in later figures can be converted to corresponding voltage 
 quantities using Fig. 2d.}

\section{Energy landscape of two dipole coupled nanomagnets (NM1 and NM2) and their magnetization dynamics}

We will assume that NM2 has major axis {\it a} = 105 nm, 
minor axis {\it b} = 95 nm and thickness {\it t} = 6 nm, to
 ensure a 32 $kT$ in-plane shape anisotropy energy barrier at room temperature.
 The numerical results in this paper depend on this value. 
Any combination of {\it a, b} and {\it t} that gives a barrier height of 32 $kT$  will yield similar results. In choosing NM2's dimensions, we only have to 
ensure that it always contains a  single domain and therefore can be 
described by macrospin dynamics
\cite{PhysRevLett.83.1042}.

The temporal evolution of the magnetization in any single-domain NM, under the influence of an 
effective magnetic field $\vec{H}_{eff}(t)$, is described by the Landau-Lifshitz (LL) equation \cite{dynamic}, 
\begin{equation}
\frac{d\vec{M}(t)}{dt} = -\gamma\vec{M}(t) \times \vec{H}_{eff}(t) - \frac{\alpha\gamma}{M_S}[\vec{M}(t) \times 
(\vec{M}(t) \times \vec{H}_{eff}(t))],
\end{equation}
where $\alpha$ is the Gilbert damping constant, $\gamma$ is the gyromagnetic ratio and $M_S$ is the saturation 
magnetization. The quantity $\vec{H}_{eff}(t)$ is the effective magnetic field acting on the magnetization
due to shape anisotropy, strain and dipole interaction. It 
is the gradient of the magnet's total potential energy with respect to the magnetization vector:
\begin{equation}
\vec{H}_{eff}(t) = -\frac{1}{\mu_0 \Omega}\frac{\partial E(t)}{\partial \vec{M}(t)},
\end{equation}
where $\mu_0$ is the vacuum permeability and $\Omega$ is the magnet's volume. 

The potential energy of an NM is given by
\begin{equation}
E= E_{shape}+E_{stress}+E_{dipole}
\end{equation}
where $E_{shape}$ is the shape anisotropy energy due to the elliptical shape of the  NM, $E_{stress}$ is the stress 
anisotropy energy caused by the stress and  $E_{dipole}$ is the dipole$-$dipole interaction energy 
between the NMs. We ignore any magnetocrystalline anisotropy energy since the 
magnets are assumed to be amorphous. The shape anisotropy energy of the $i^{\rm th}$ NM in spherical coordinates, $\theta_i$ 
and $\phi_i$, is
\begin{equation}
\begin{split}
E_{shape}= \frac{\mu_0 M^2_S \Omega}{2}(Nd_{xx}sin^{2}\theta_i cos^{2}\phi_i &+Nd_{yy}sin^{2}\theta_i sin^{2}\phi_i \\
&+Nd_{zz}cos^{2}\theta_i)
\end{split}
\end{equation}
where $Nd_{xx}$ , $Nd_{yy}$ and $Nd_{zz}$ are respectively the demagnetization
factors along the x-, y- and z-directions and are dependent on $a$, $b$ and $t$.  The equations to calculate the demagnetization factors can be found in \cite{Chikazumi,0957-4484-23-10-105201}. The stress anisotropy energy in the 
$i^{\rm th}$ element 
due to a
stress applied along its major axis is
\begin{equation}
\begin{split}
E_{stress}= -\frac{3}{2}\lambda\sigma\Omega sin^{2}\theta_i sin^{2}\phi_i 
\end{split}
\end{equation} 
where $(3/2)\lambda$ is the is the saturation magnetostriction and $\sigma$ is the applied stress. 
The quantity $\sigma$ is negative for 
compression and positive for tension\cite{0957-4484-23-10-105201}. The dipole-dipole interaction energy between the $i$-th 
and $j$-th NMs is 
\begin{equation}
\begin{split}
E_{dipole}= &\frac{\mu_0 M_S^2 \Omega_i \Omega_j}{4\pi d^3}[-2(sin\theta_i cos\phi_i )(sin\theta_j cos\phi_j )\\
&+ (sin\theta_i sin\phi_i  )(sin\theta_j  sin\phi_j) + cos\theta_i cos\theta_j  ]
\end{split}
\end{equation} 
where $d$ is the separation between their centers. For this study, we take d to be 150nm. We have assumed that the magnetostriction  $(3/2)\lambda_S = 9\times 10^{−4}$ \cite{1059598} and saturation magnetization $M_S = 0.8\times 10^6 Am^{-1}$ \cite{PSSA:PSSA195}. The Gilbert damping constant for Terfenol-D is $\alpha = 0.1$\cite{0022-3727-41-16-164016}. 

In the macrospin limit, the NMs were assumed to have uniform magnetization. To show 
that the macrospin limit is a good approximation, the switching dynamics were compared with detailed 
3D micromagnetic simulations. Two micromagnetic packages were used: The Object Oriented MicroMagnetic 
Framework (OOMMF)\cite{OOMMF} and M$^3$\cite{M3}. Details of the comparison is in Appendix I.

The critical stress is the amount of stress needed to dislodge an NM's magnetization 
 from a stable orientation along the easy axis. This quantity 
 depends on the in-plane shape anisotropy energy barrier 
and the magnetization orientation  of its neighbors
that determine the dipole interaction energy. The critical stresses required to rotate the magnetization of 
 NM2 in isolation (no neighbors), 
 in the $\uparrow\uparrow$ configuration with NM1, and in the $\uparrow\downarrow$ configuration with NM1 are, respectively:

\begin{equation}
\sigma^C_{single~NM}=\frac{\mu_0}{2\lambda} M^2_S [Nd_{zz}-Nd_{yy}]
\end{equation} 

\begin{equation}
\sigma^C_{\uparrow\uparrow~with~NM1}=\frac{\mu_0}{2\lambda} M^2_S [Nd_{zz}-Nd_{yy}-\frac{\Omega_{NM1}}{2\pi d^3}]
\end{equation} 

\begin{equation}
\sigma^C_{\uparrow\downarrow~with~NM1}=\frac{\mu_0}{2\lambda} M^2_S [Nd_{zz}-Nd_{yy}+\frac{\Omega_{NM1}}{2\pi d^3}]
\end{equation}
 
These quantities are 3.14MPa, 2.72MPa and 3.56MPa, respectively, for $d$ = 150 nm.

\begin{figure}[t]
\centerline{\epsfig{figure=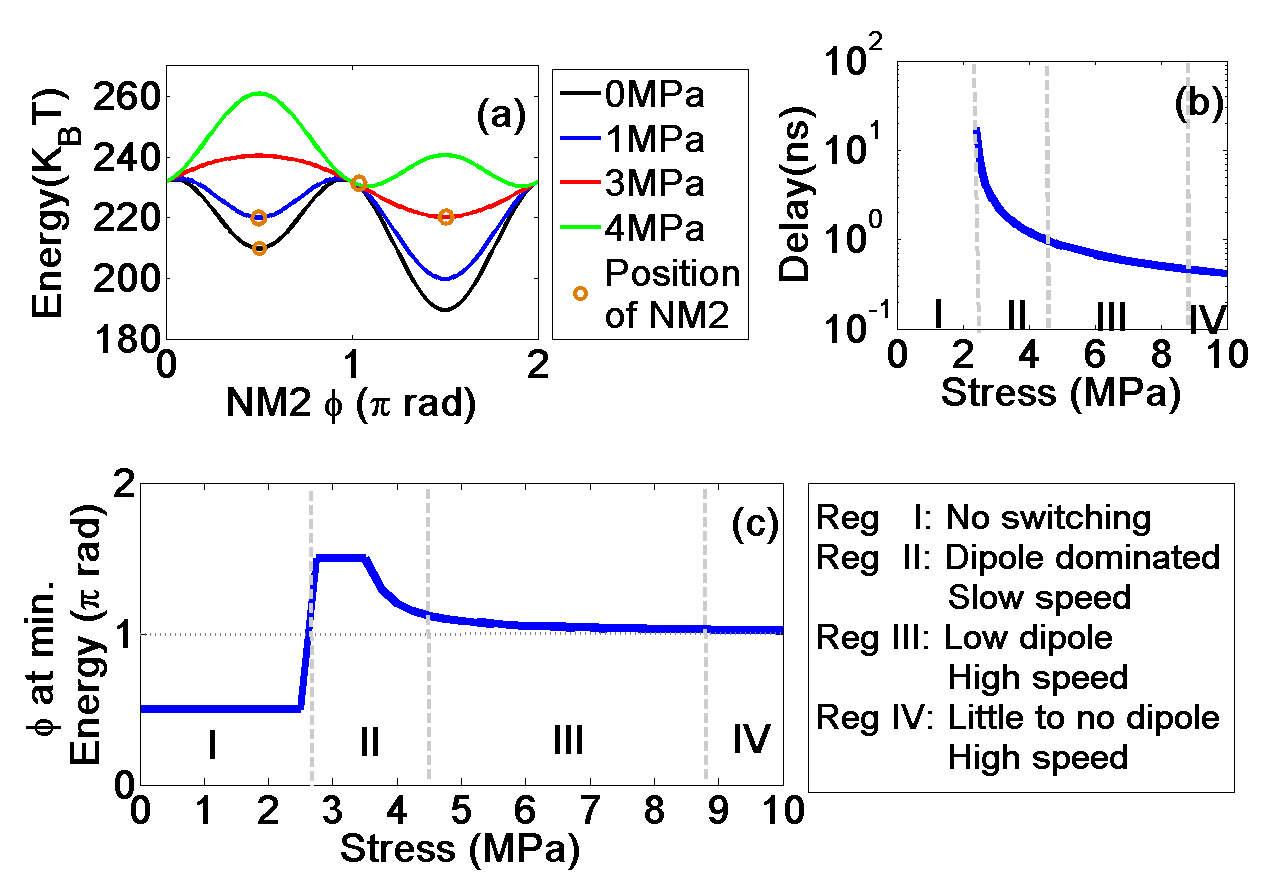,width=3.5in,height=2in}}
\caption{(a) The potential energy landscape ($E$ as a function of $\phi$) and the stable location (at 0 K)
of the magnetization 
vector 
of NM2 (depicted with circles) at different stresses. The stable location 
will be always at the local energy minimum closest to the starting location.
(b). Switching delay as a function of stress. This figure also demarcates the stressing 
regions defined in (c). (c) The stable location of magnetization vector corresponding to the local 
energy minima (depicted by the azimuthal angle $\phi$ of the vector) for different stresses. In Region I, 
corresponding to sub-critical stresses, the energy minimum is at  $\phi = \pi/2$. In the interval 2.75 to 4.5 MPa 
that straddles the critical stress (Region II), the minimum energy location $\phi$ is close to $3\pi/2$.  
In regions III and IV, the applied stress is super-critical and the  minimum energy location moves to $\phi=\pi$
which is the in-plane hard axis.}
 \label{band_structure}
\end{figure}

Fig. 3a shows the energy landscape and the stable (minimum energy) location of the magnetization vector of NM2 at 
different stress levels. 
At sub-critical stresses, the initial orientation of the magnetization ($\phi=\pi/2$) remains a local energy minimum 
so that the magnetization is
trapped there and does not rotate. Close to the critical stress level of 3 MPa, the applied stress begins to 
cancel out the in-plane shape anisotropy energy barrier and the dipole 
interaction with NM1 dominates the energy landscape
(we referred to this as the {\it{dipole-dominated regime}}). The energy minimum therefore moves to $\phi=3\pi/2$ since dipole 
interaction favors anti-ferromagnetic ordering. As a result, the magnetization will rotate to the
 desired location 
(i.e. flip as desired), but
because the stress is weak, rotation is slow and takes around 50 ns (Fig. 3b). At higher stress levels, the energy minimum 
shifts to the in-plane hard axis ($\phi=\pi$) since the stress anisotropy energy becomes overwhelmingly dominant.  The high stress level speeds up the rotation. In Fig. 3c, we define 
three regions based on the location of the energy minimum. In Region I, for sub-critical stresses, 
the energy minimum is at  $\pi/2$ and the  magnetization does not switch at all. For the dipole dominated regime between 
2.75 and 4.5 MPa (Region II), the minimum energy $\phi$ is close to $3\pi/2$ so that the 
magnetization ultimately flips, albeit slowly since the stress is weak.  In Regions III and IV, the stress starts to 
dominate the energy landscape and the minimum energy location is at 
$\phi = \pi$. These different regions are associated with different switching error probabilities.

\begin{figure}
\centerline{\epsfig{figure=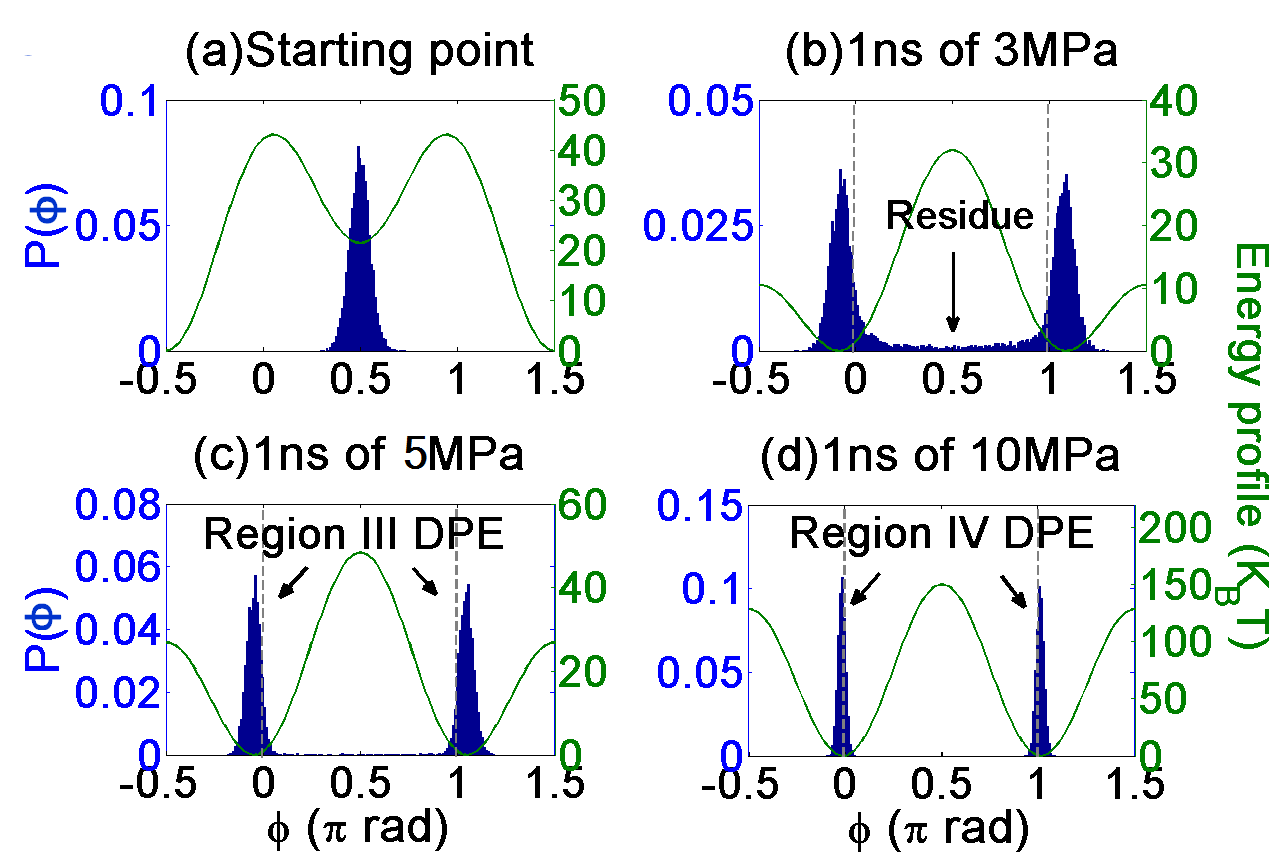,width=3.5in,height=2in}}
\caption{(a) Distributions of the azimuthal angle $\phi$ (for $\theta = \pi/2$) of the  magnetization vector of NM2 due to 
thermal fluctuations during step A at room temperature. The peak of the distribution is at the stagnation point 
$\phi=\pi/2$ showing that the most likely initial location of the magnetization is unfortunately the stagnation point. 
This is expected since the stagnation point is a local energy minimum in the unstressed magnet. 
The solid curves in each case show the potential energy profile in NM2, i.e. potential as a function of $\phi$
for $\theta = \pi/2$. (b) A super-critical 
stress of 3 MPa turned on abruptly for 1 ns is not enough to kick all switching trajectories out of the stagnation point 
and toward the hard axis. The peaks of the distribution do shift close to $\phi$ = 0 or $\pi$ because the stress is 
super-critical and the magnetization is likely to rotate to the in-plane hard axis, but the distribution at 
$\phi = \pi/2$ is not completely depleted showing that there is a significant probability of the magnetization
vector not rotating successfully to the hard axis and remaining stuck at the easy axis (failure). The peaks are not 
exactly at $\phi$ = 0 or $\pi$, but instead are slightly displaced from these locations because of the effect of dipole 
coupling which is not completely overwhelmed by stress. (c) A stress of 5 MPa turned on abruptly for 1 ns is strong enough 
to kick most switching trajectories out of the stagnation point and thus nearly deplete the distribution at $\phi = \pi/2$.
 The peak of the switched distribution is less displaced from the hard axes at $\phi=0~or~\pi$ than in the previous case
because the stress is stronger but still not strong enough to completely overwhelm the dipole coupling effect. (d) A very
strong stress 
of 10 MPa turned on abruptly for 1 ns completely depletes the distribution at $\phi = \pi/2$
showing that nearly every switching trajectory has been kicked out of the stagnation point.
This time, the peaks are almost exactly at $\phi=0~or~\pi$, i.e. the in-plane hard axes, since the strong stress almost 
completely overwhelms the dipole effect.}    \label{band_structure}
\end{figure}

\section{Information transmission between two NMs taking thermal perturbations into account}

Fig. 4a shows the room-temperature distribution of the azimuthal angle $\phi$ of 
the magnetization vector in NM2 during step A when it is in  parallel configuration with 
NM1 at a distance of 150 nm. The thermal perturbations on the magnetization 
vector of NM2 are modeled by a Langevin random 
field $\vec{H}_{L}(t)$  that can be added to the effective magnetic field term in the 
LL equation. This field $\vec{H}_{L}(t)$ is related to the temperature $T$ by, 
\begin{equation}
\vec{H}_{L}(t) =\sqrt{\frac{2\alpha k T}{\mu_0 \gamma \Omega M_S \delta t }}\vec{G}(t)
\end{equation}
where $\vec{G}(t)$ is a white noise whose three components in
Cartesian coordinates are Gaussians with zero mean and unit standard deviation \cite{sun2}. 
The quantity $\delta t$ is inversely proportional to the 
attempt frequency of thermal noise to flip magnetization and is
the simulation time-step used to solve the coupled Landau-Lifshitz-Gilbert equation 
(describing the coupled dynamics of NM1 and NM2) numerically \cite{brown}.  
The peak of the distribution is at the stagnation point $\phi=\pi/2$ where $\delta \phi/\delta t=0$ for any applied stress.

\begin{figure}
\centerline{\epsfig{figure=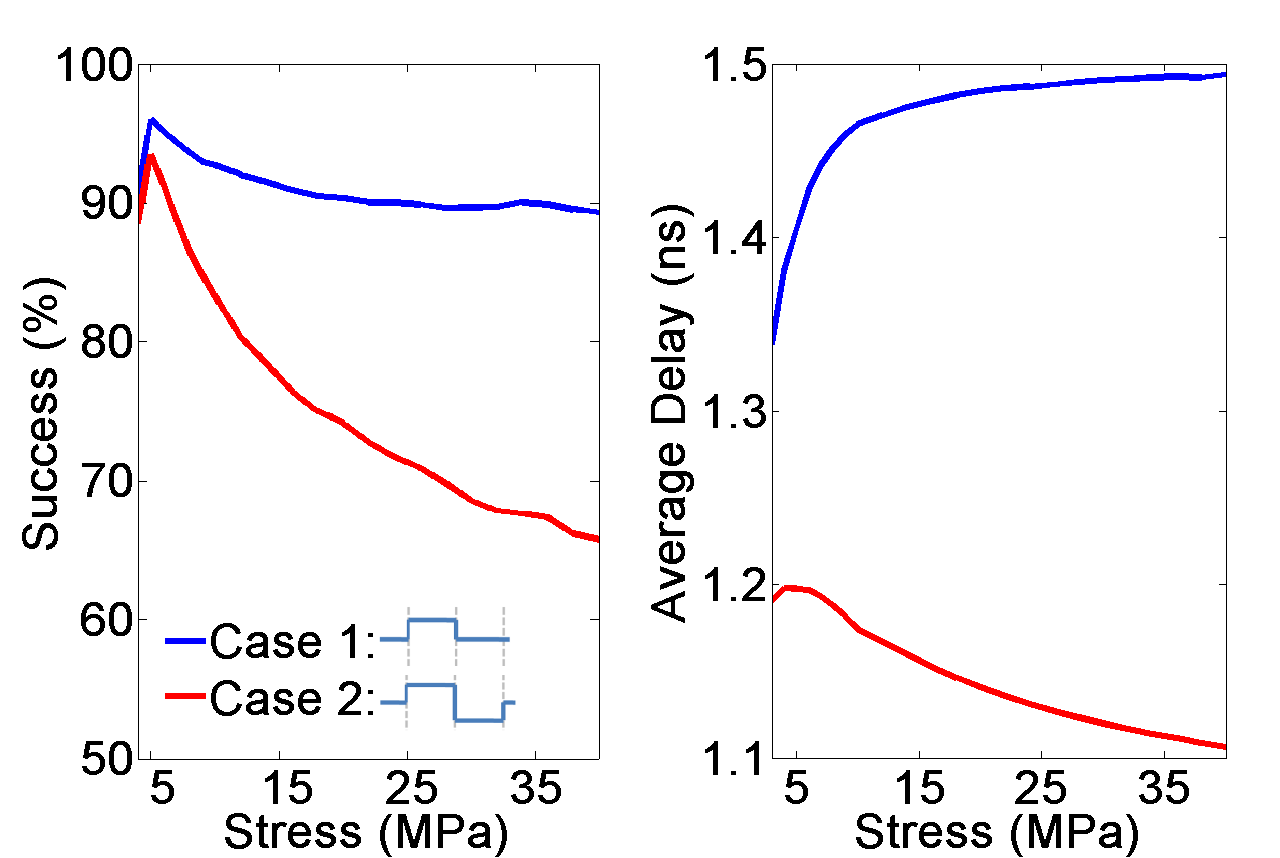,width=3.5in,height=1.8in}}
\caption{The probability of successful switching of 
NM2 after NM1 is flipped (left panel) and the 
thermally averaged switching delay (right panel) as a function of stress amplitude for the stressing profiles or
pulse shapes in Cases 1 and 2. The results are for room temperature (300 K).
In both cases, the success rate increases with 
stress 
at smaller stress levels, peaks around 5 MPa and then slowly falls off as stress is increased. The fall-off rate 
is much larger in case 2 than in case 1. The thermally averaged switching delay in case 1 is around 1.38 ns at the 
optimum stress of 4 MPa, while in case 2, it is 1.21 ns because stress reversal in case 2 speeds up
switching.} \label{band_structure}
\end{figure}

\begin{figure}
\centerline{\epsfig{figure=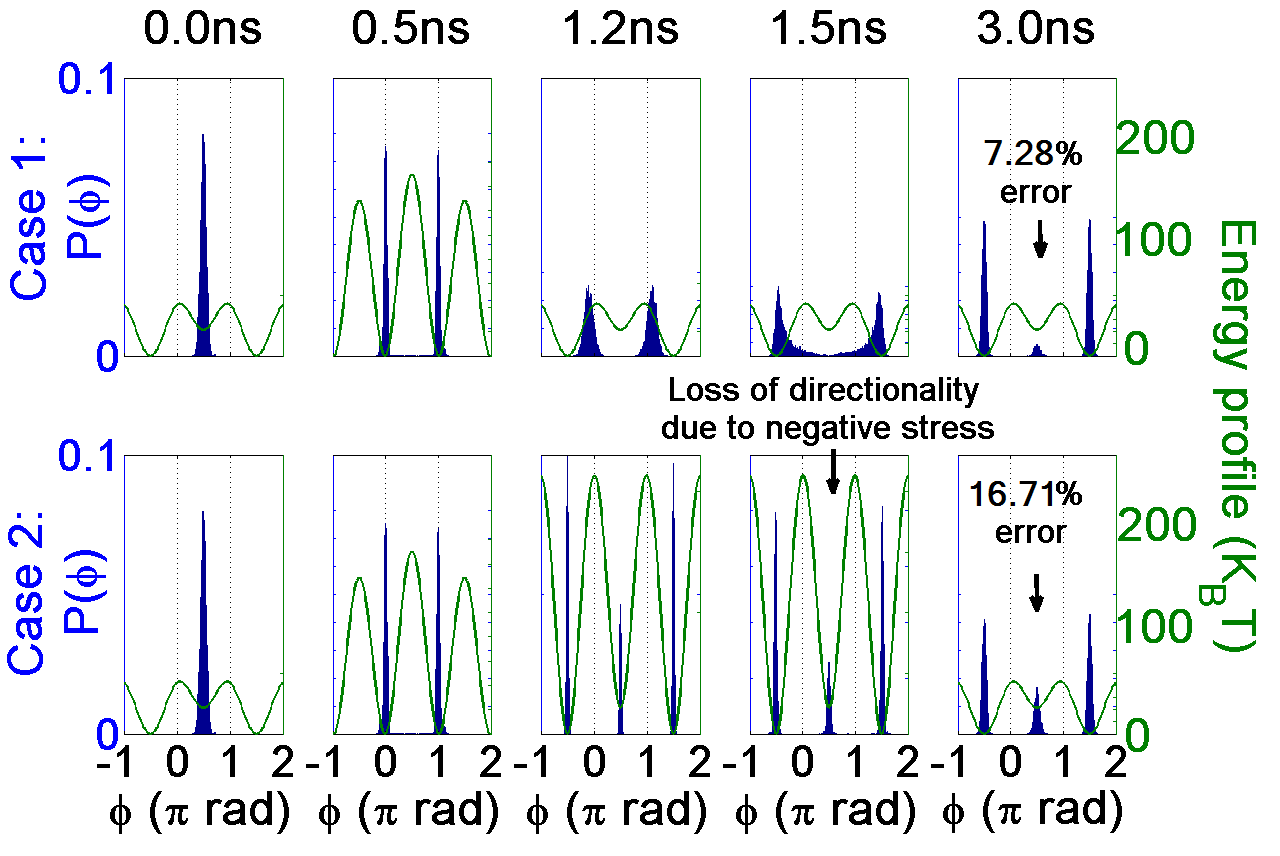,width=3.5in,height=2.2in}}
\caption{The room temperature distributions of the 
azimuthal angle of the magnetization vector and the potential profiles in NM2
at different instants of time for the 
pulse shapes of cases 1 and 2. The stress amplitude is 10 MPa. 
While the negative stress in case 2 speeds up the switching process, it 
weakens the dipole coupling with NM1 and therefore causes a higher error rate.} \label{band_structure}
\end{figure}

\section{Switching reliability}

In this section, we study switching error probability as a function of pulse {\it shape} 
of the Bennett clock (stress versus time profiles) at room temperature.  We also discuss trade off between error resilience and switching
speed.

\subsection{Case 1:  abrupt stress application and removal}
We consider the situation when the stress profile is a rectangular pulse
turned on and off abruptly
with infinite ramp rate (see case 1 in Fig. 2b). The pulse has a fixed width of 1 ns (legend of Fig. 5).  Clearly, 
a pulse amplitude of 
3 MPa is not sufficient to kick the magnetization vector out of the stagnation point with
absolute certainty at the end of step B because of the appreciable residue around the starting orientation 
 in the probability 
distribution of the magnetization (Fig 4b). This residue results in a low success rate.  At a higher stress amplitude 
of 5 MPa  
(Fig. 4c, region III), the magnetization vector is kicked out of the starting orientation 
with high probability (near vanished residue). This stress is still low enough to allow dipole 
interaction to flip NM2 and steer the system to the anti-parallel configuration
($\phi = 3 \pi/2$) with relatively high certainty. If we operate in this stress regime, then
we will obtain a high success rate for switching to the correct 
anti-parallel configuration. If now the stress is further increased 
(region IV, Fig 4d), the influence of dipole interaction is diminished and the distributions become more symmetric about 
the hard axis. This means that upon releasing the stress, the 2-NM system has only slightly higher 
chances of going to the ground state (anti-parallel configuration and successful switching) 
than the metastable state  (parallel configuration and switching failure).  This explains why the success 
rate has a non-monotonic 
stress dependence and begins to decrease 
with increasing stress beyond an optimum stress (Fig. 5).

The highest success rate achieved with a 1 ns wide rectangular pulse is 95.97\%
at room temperature. It is achieved when the stress is 5 MPa.  The thermally averaged 
switching delay at this 
stress is 1.38 ns. The last quantity is calculated 
by averaging over the delays associated with the successful switching trajectories
among 10,000  trajectories whose starting points are chosen from the 
distributions of the initial magnetization orientation. The switching trajectories are computed by solving the stochastic 
LL equation (Equation (1) where ${\vec H}_{eff}(t)$ is replaced with  ${\vec H}_{eff}(t) + {\vec H}_L(t)$). 

\subsection{Case 2:  abrupt stress application, reversal  and removal}
In order to switch faster, we can reverse the stress after 1 ns, so that a negative stress 
is applied during step C (legend of Fig.5). As expected, the success probability still peaks at the optimum stress of 5
MPa, but this probability is smaller than in Case 1. The success
probability at high stresses also drops off much faster with 
increasing stress compared to Case 1. All this happens because negative stress actually raises the 
energy barrier between the two stable orientations along the easy axis instead of lowering this barrier. 
Therefore, in step C, the {\it difference} caused by dipole interaction between the energies of the 
two stable states of NM2 becomes  a {\it smaller} fraction of the barrier height. 
As a result, the preference for the ground state of the inverter over the metastable state decreases. 
We call this loss of `dipole directionality' since the dipole-induced preference for 
the anti-parallel ordering is eroded, resulting in
a higher error rate. The average switching delay at the optimum stress however decreases and is now
1.2 ns. Case 2 therefore results in faster switching but a higher error rate compared to Case 1.

\begin{figure}
\centerline{\epsfig{figure=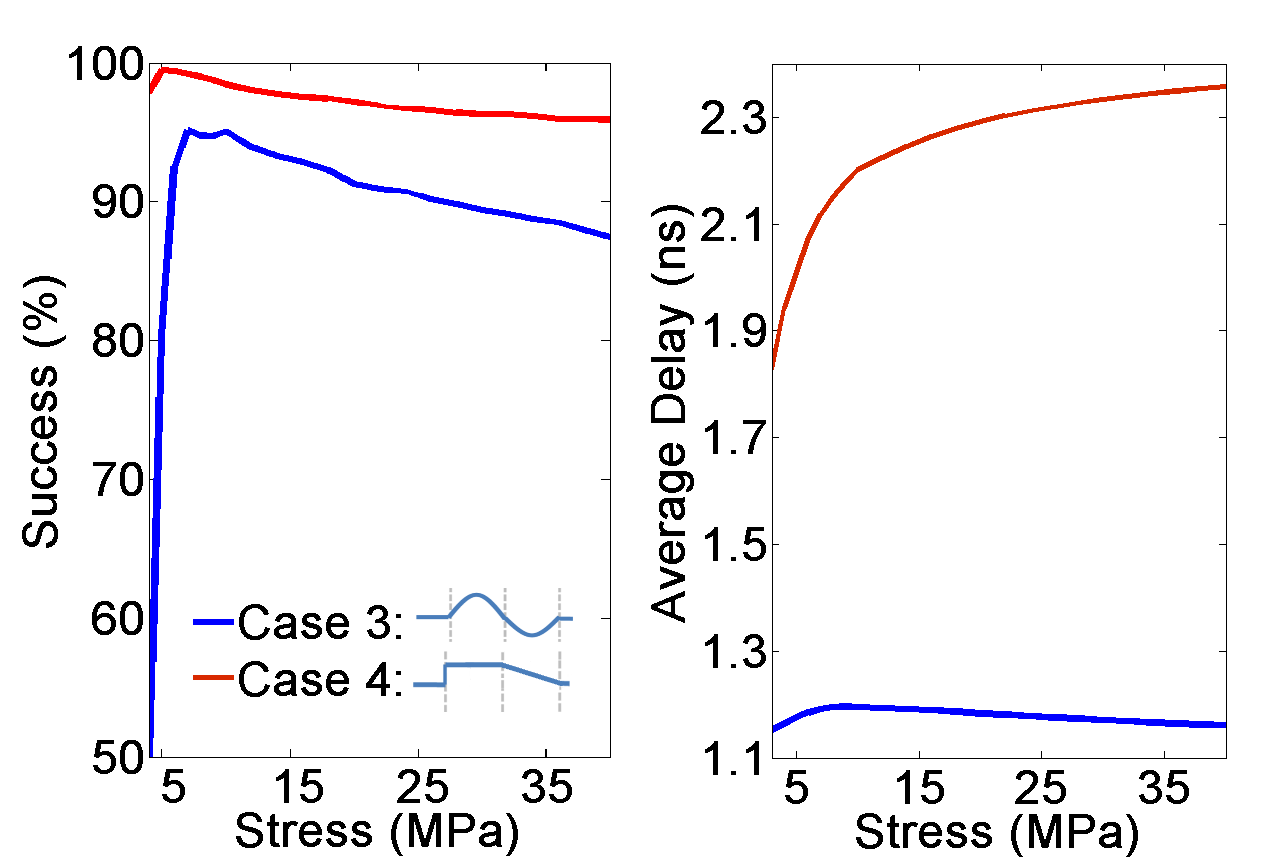,width=3.5in,height=1.8in}}
\caption{Probability of successful switching of NM2 after NM1 is flipped (left panel)
and thermally averaged switching delay (right panel) as a function of stress amplitude
for the pulse shapes in Cases 3 and 4 at room temperature. In Case 3, the overall success rate is lower since the sinusoidal 
stress during 
step B is not enough to kick the magnetization out of the stagnation point with near certainty. Furthermore,
 the negative stress during step C increases the energy barrier between the 
two stable states along the easy axis (local and global energy minima)
and 
diminishes the influence of dipole coupling responsible for the energy difference between the 
two minima. Reduced influence of dipole coupling impairs successful switching.
The thermally averaged switching delay in Case 3 is 1.13 ns. Case 4 yields better success probability but 
has a longer thermally averaged delay of 2.02 ns because stress is never reversed to 
speed up switching.}  \label{band_structure}
\end{figure}

\begin{figure}
\centerline{\epsfig{figure=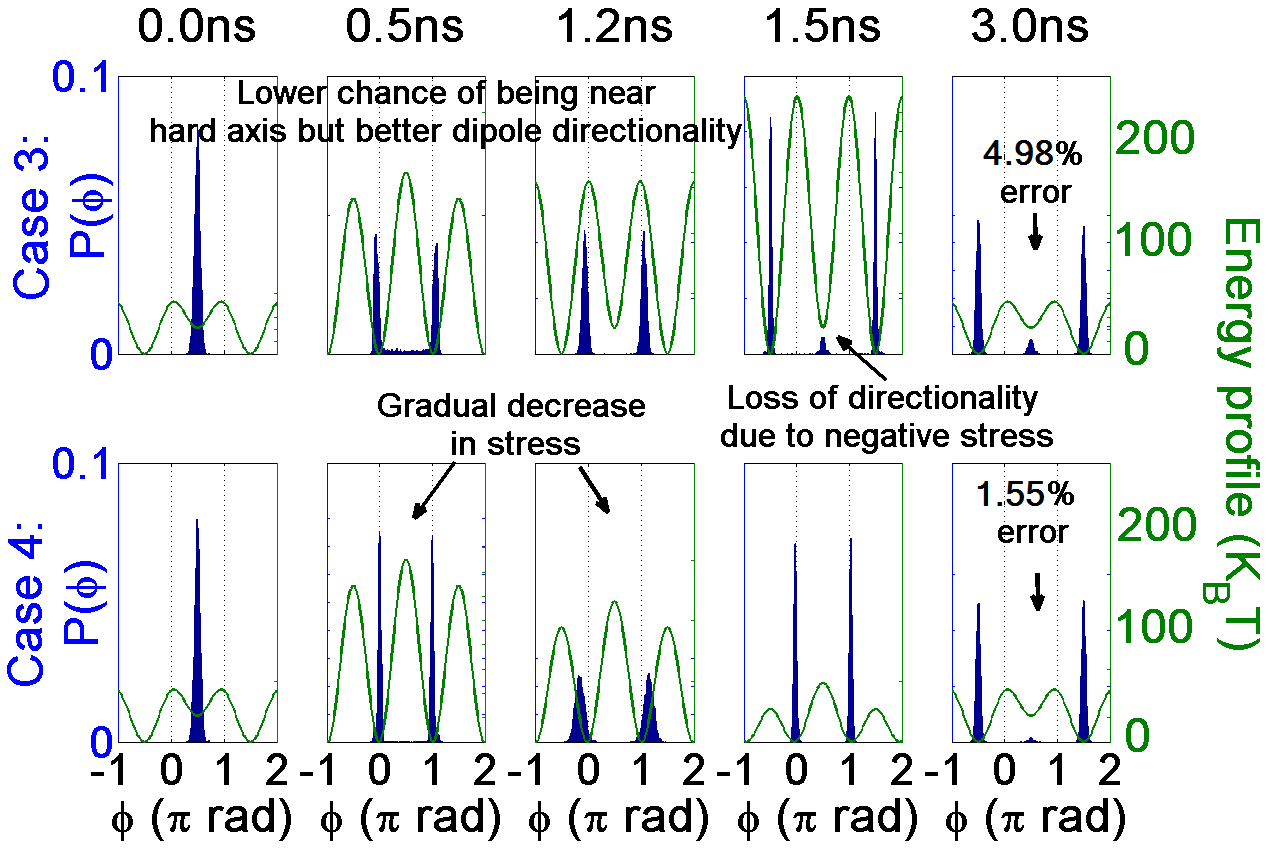,width=3.5in,height=2.2in}}
\caption{The room temperature distributions of the 
azimuthal angle of the magnetization vector and the potential profiles in NM2
at different instants of time for the 
pulse shapes of cases 3 and 4. The stress amplitude is 10MPa. 
While the stress reversal in case 3 speeds up the switching process, it 
weakens the dipole coupling with NM1 and therefore causes a higher error rate
of 4.98\% compared to 1.551\%  in case-4 at 10 MPa stress. } \label{band_structure}
\end{figure}

\subsection{Cases 3 and 4: sinusoidal stress and tapered stress removal}
Case 3 pertains to a sinusoidal stress profile with a period of 2 ns (Fig. 7).
Stress reversal during the negative cycle of the sinusoid
 switches the magnetization of NM2 faster, but the overall success rate is low as the sinusoidal stress during step B 
is not strong enough to kick the magnetization out of the stagnation point 
with high probability (Fig. 7). Furthermore, the negative stress during step C diminishes the role of dipole coupling 
as in Case 2. The thermally averaged switching delay at the stress level where 
success probability peaks is 1.2 ns (Fig.7). 

 Case 4 pertains to a linear ramp for stress withdrawal that we term tapered stress removal.
After keeping the stress on for 1 ns, it is gradually removed in 1 ns instead of abruptly (with a constant ramp rate), 
achieving higher reliability than Cases 1, 2, and 3.  The slow removal of stress allows the magnetization switching 
to operate in the dipole dominated stress region for a longer amount of time (Fig. 8). The thermally averaged 
switching delay is 2.02 ns at the stress level where the success probability peaks
(Fig.7). For both cases 3 and 4, the success probability decreases with increasing stress beyond the 
optimum stress value because 
we end up spending less time in regions II and III as stress is increased. 

\begin{figure}[t]
\centerline{\epsfig{figure=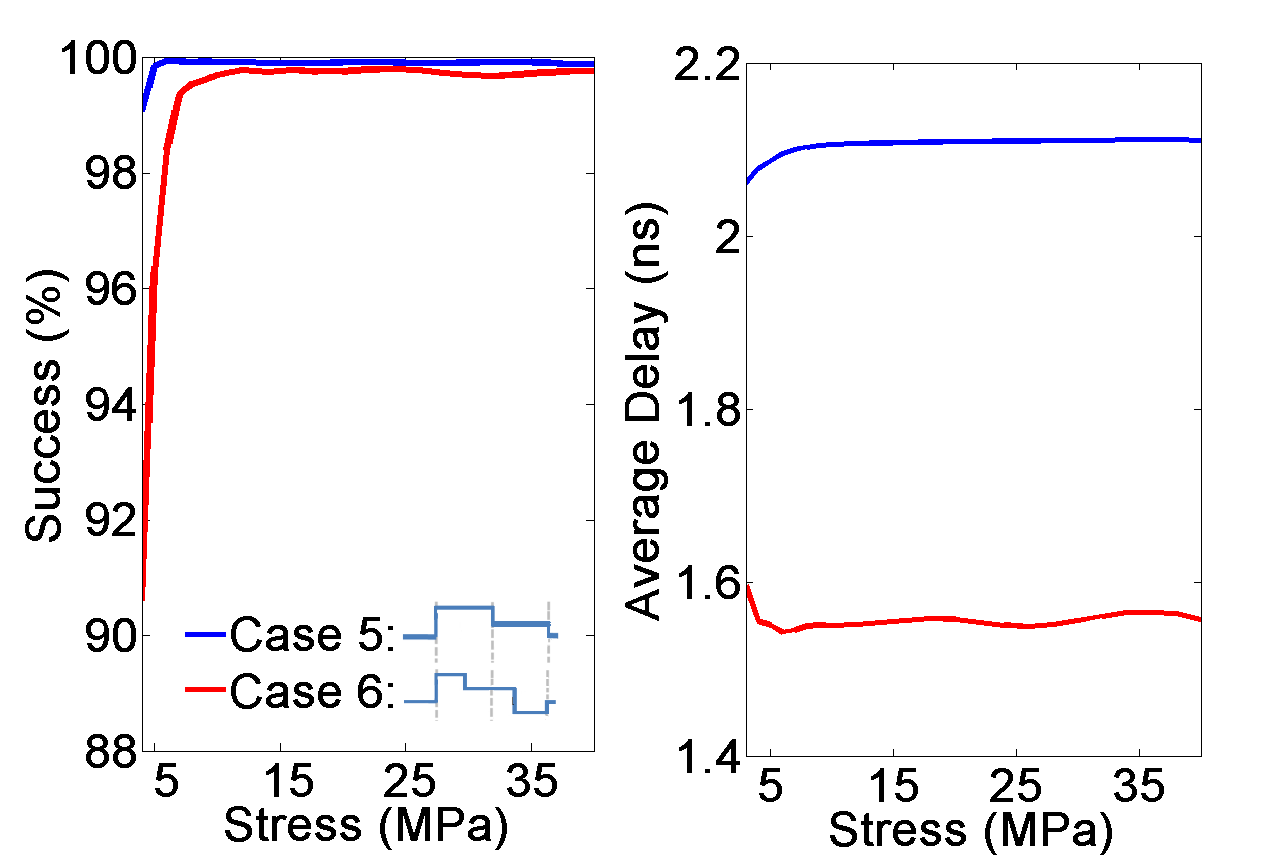,width=3.5in,height=1.8in}}
\caption{Probability of successful switching of NM2 after NM1 is flipped (left panel)
and thermally averaged switching delay (right panel) as a function of the initial stress 
for the pulse shapes in Cases 5 and 6 at room temperature. For these
pulse shapes, the success probability increases monotonically with 
stress unlike in the previous four cases. In Case 6, the success rate is lower because 
of a period of stress reversal which diminishes the role of dipole coupling but makes the switching faster.
The thermally averaged switching delays at room temperature in Cases 5 and 6 are1.8 ns and 1.33 ns, respectively, at stress
levels of 10 MPa.} \label{band_structure}
\end{figure}

\begin{figure}[t]
\centerline{\epsfig{figure=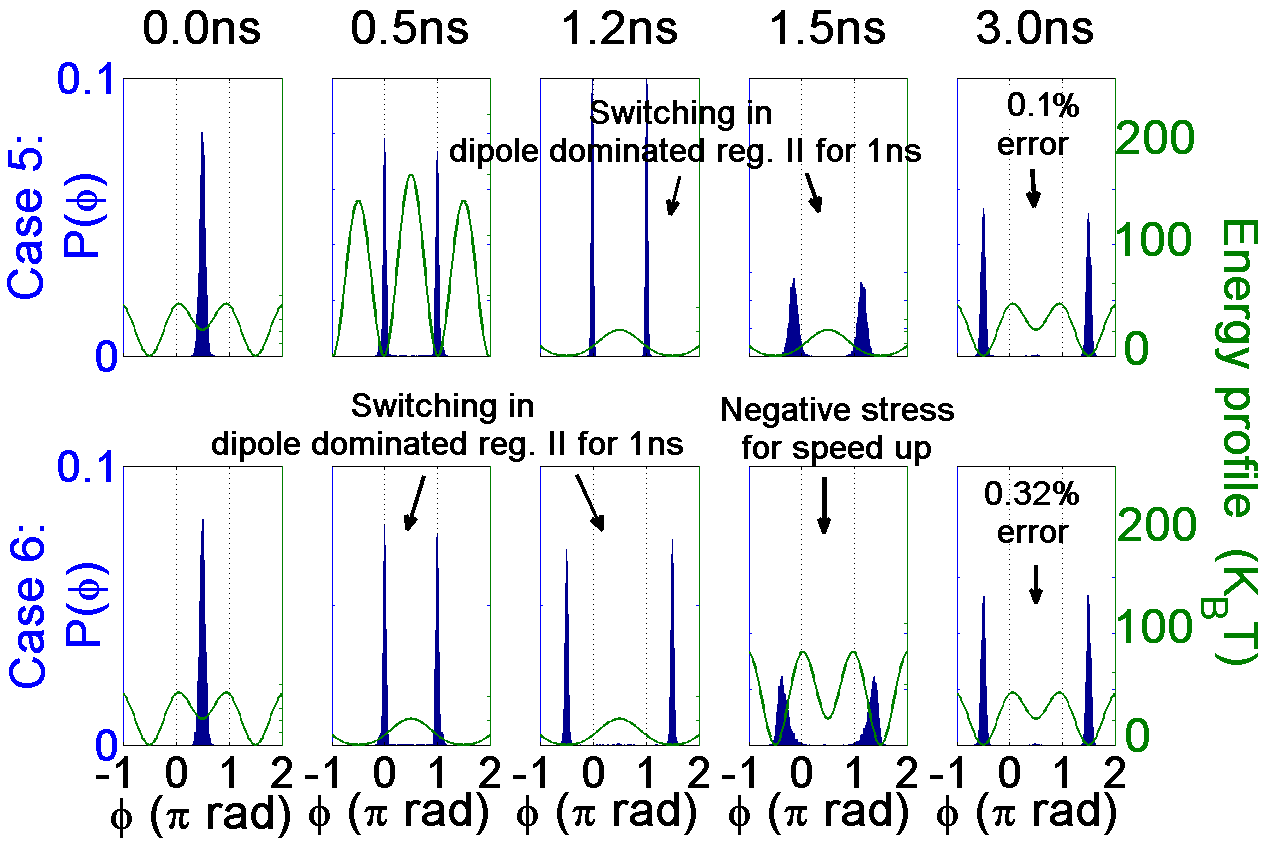,width=3.5in,height=2.2in}}
\caption{The room temperature distributions of the 
azimuthal angle of the magnetization vector and the potential profiles in NM2
at different instants of time for the 
pulse shapes of cases 5 and 6. The stress amplitude is 10MPa. 
Operation in the nearly dipole dominated region for 1 ns has drastically reduced 
the error probability. Switching is faster in Case 6 because of a period 
of stress reversal but at the cost of more than twice the error probability
at 10 MPa.} \label{band_structure}
\end{figure}

\begin{figure}[t]
\centerline{\epsfig{figure=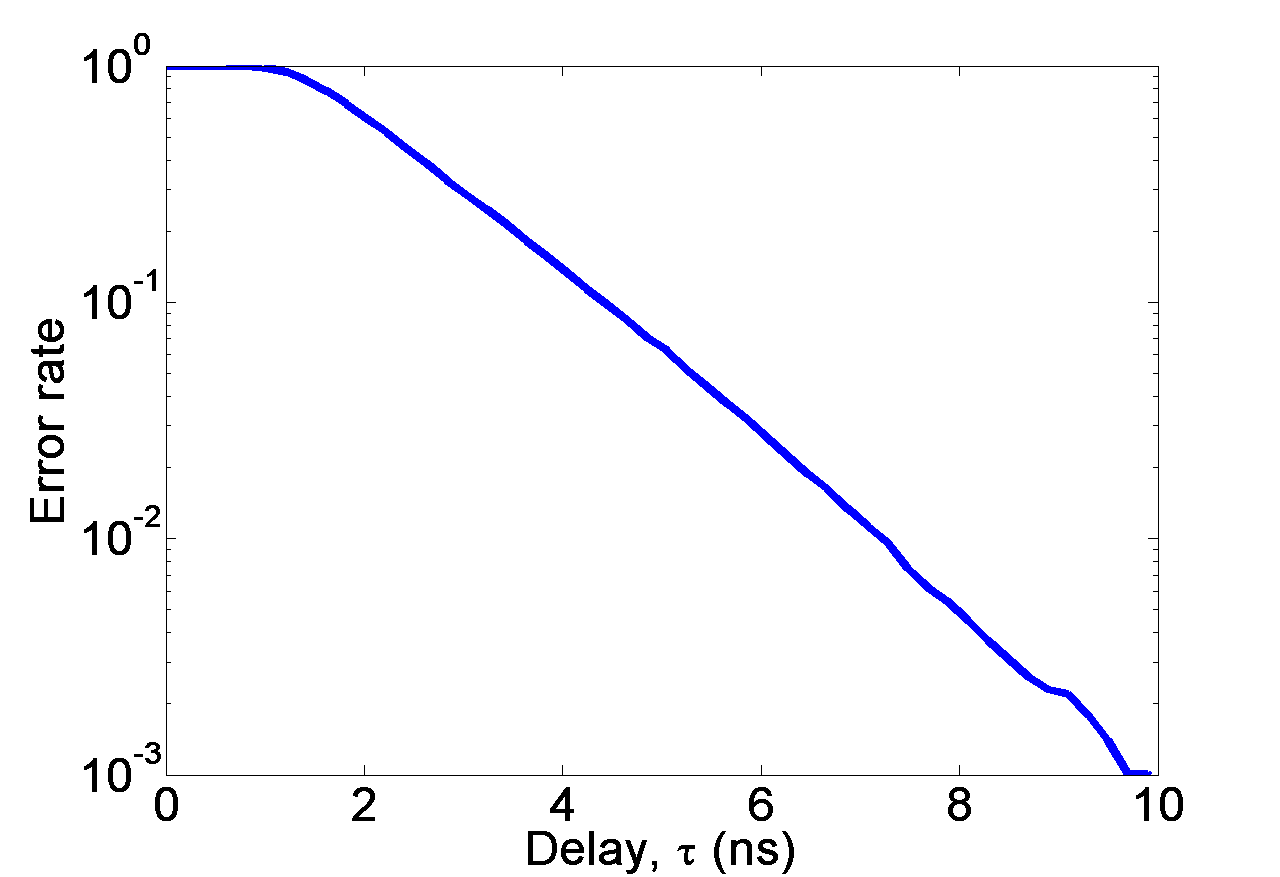,width=3.5in,height=1.8in}}
\caption{Very low error rate can be achieved 
by abruptly turning 10 MPa of  stress on NM2, decreasing 
it to 3 MPa abruptly after 1 ns, and then holding it for a long time. 
Error rate of 10$^{-3}$ is achieved by holding the stress for  10 ns at room temperature. } \label{band_structure}
\end{figure}

\subsection{Switching efficiently using region II - the dipole dominated regime}

 Cases 5 and 6 pertain to complex stressing profiles or pulse shapes shown in the 
legend of Fig. 9. Initially, in Case 5, a high stress 
is applied abruptly to NM2 to kick the magnetization out of the stagnation 
point. 
Then the stress is reduced abruptly to 3 MPA for 1 ns and finally withdrawn abruptly. 
During the last 1 ns (Fig. 10), the switch operates in region II where the dipole interaction dominated energy 
landscape nudges the magnetization to the correct final orientation to complete switching successfully. 
The success probability is much higher (99.9\% for an initial stress of 10 MPa with a thermally averaged delay of 
1.8 ns at room temperature).
   
 In Case 6, a high stress is turned on abruptly, held for 0.5 ns, then reduced abruptly to 3 MPa, held for 1 ns,
  reversed abruptly to -3 MPa, held for 0.5 ns, and finally withdrawn abruptly.
The stress reversal, as always, accomplishes faster switching 
(thermally averaged delay of 1.33ns) but at the cost of a slightly lower 
success probability of 99.68\% for 10 MPa initial stress at room temperature. 

Unfortunately, even the best case scenario (case 5) with a room temperature
success probability of 99.92\% may not be 
good enough for contemporary logic which has stringent requirements on switching error 
probability. If we modify the stressing profile of Case 5
by continuing to stress NM2 at 2MPa beyond the 1 ns duration, a much lower error rate can be achieved (Fig. 11). 
However, this low error rate comes at the expense of a very long switching delay of 10 ns for an error probability of 
10$^{-3}$. Note that in this range, the error probability falls off nearly exponentially with increasing delay. 
If this trend could be extrapolated to much longer delays, then we could reach an error probability of 
$10^{-8}$ at delays of $\sim$30 ns, provide the out-of-plane magnetization effects do not begin to dominate first. 
That error rate may be tolerable.

\textcolor{black}{Another strategy to reduce error rate (without any pulse shaping) is to increase the shape anisotropy energy 
barrier of the
magnets (make the ellipses more eccentric) which will allow us to increase the dipole coupling strength without 
inducing ferromagnetic ordering. The increased dipole coupling reduces error probability, but the stress needed
to switch the magnets by overcoming the shape anisotropy energy barrier is now also larger, resulting in increased energy
dissipation.  Since SML is attractive for its low energy, it is imperative to keep the dissipation as small as 
possible,
which is why we have not explored this case. Nevertheless, it should be kept in mind that there are two ways to 
reduce error probability: (1) by increasing switching delay (which we have discussed), or (2) by increasing 
energy dissipation (which we have alluded to but not discussed in detail). In the end, there is always a trade off
between error probability and energy-delay product; better error-resilience can be purchased with higher 
energy-delay product.}

\begin{table}[t]
\caption{Error rate and delay comparison}
\begin{center}
\begin{tabular}{|c|c|c|c|}
\hline
Case & Stress where   & Peak success   & Average switching\\
&  success rate  & probability (\%) & delay at peak(ns)\\
& peaks (MPa) && \\
\hline
1 & 5& 95.97 &1.38\\
\hline
2 &5&93.45 &1.21\\
\hline
3 &7& 95.12 &1.13\\
\hline
4 & 5& 99.49 &2.02\\
\hline
5 & 6& 99.92&1.82\\
\hline
6 & 12& 99.77&1.33\\
\hline

\hline

\end{tabular}
\end{center}

\end{table}

\section{Conclusion}

 In this work, we used physical insights gained from the energy profiles of
stressed magnets to devise various stressing profiles to reduce error rates in SML.
Error rates small enough for logic may be achieved, but at very long switching delays.
Thus, one disadvantage of dipole-coupled NML  seems to be that it is very slow (clock speed should be 
few tens of MHz for reliable operation with acceptable error probabilities for logic at room temperature), but the 
energy advantage 
of SML still makes it an attractive paradigm for niche applications where speed is not the main concern, but 
energy-efficiency
is. The application most suitable for SML-type technology is in medically implanted devices that need to harvest 
energy from the patient's body movements without requiring a battery to operate. This has been discussed in 
ref. \cite{roy:063108}

\setcounter{figure}{0} \renewcommand{\thefigure}{A.\arabic{figure}}

\section*{Acknowledgments}
K. Munira and Y. Xie would like to thank Claudia K.A. Mewes and Andrew Tuggle from University of Alabama for useful discussion regarding M$^3$. All authors acknowledge support from the US National Science Foundation under NEB 2020 grant ECCS-1124714 
and  the Semiconductor Research Corporation (SRC) under NRI Task 2203.001. J. A. and S. B also received support from the National Science Foundation under the SHF-Small grant CCF-1216614. J.A acknowledges support from NSF CAREER grant CCF-1253370.

\section*{Appendix I}
To study the candidacy of various stressing profile for NML, 
the NMs were assumed to have uniform magnetization and their switching dynamics 
were studied with macrospin approximation. To show that the NMs can be treated as single 
domain magnets, their  switching dynamics were compared with detailed three dimensional 
micromagnetic simulations. Two micromagnetic packages were used: The Object Oriented MicroMagnetic 
Framework(OOMMF)\cite{OOMMF} developed in National Institute of Standards and Technology (NIST) 
and M$^3$\cite{M3} in University of Alabama. In the micromagnetic simulation, the NM is divided 
into small cells and the magnetization in each cell is assumed to be uniform.  The dimension of 
the NM is $105 {nm}\times 95 {nm}\times 6 {nm}$. The cell size is $2{nm}\times2{nm}\times2{nm}$. 
Time integration of  Landau-Lifshitz-Gilbert(LLG) equation is performed on each cell with effective 
field coming from exchange coupling, shape anisotropy and external magnetic field. 
To simulate Terfenol-D nano-magnet, the following parameters were used: 
saturation magnetization $M_S = 0.8\times 10^6 Am^{-1}$, exchange stiffness 
$A=9 \times 10^{-12} J/m$ \cite{5018466}, and Gilbert damping factor $\alpha=0.1$.  
No magnetocrystalline anisotropy was taken into consideration.

\begin{figure}[t]
\centerline{\epsfig{figure=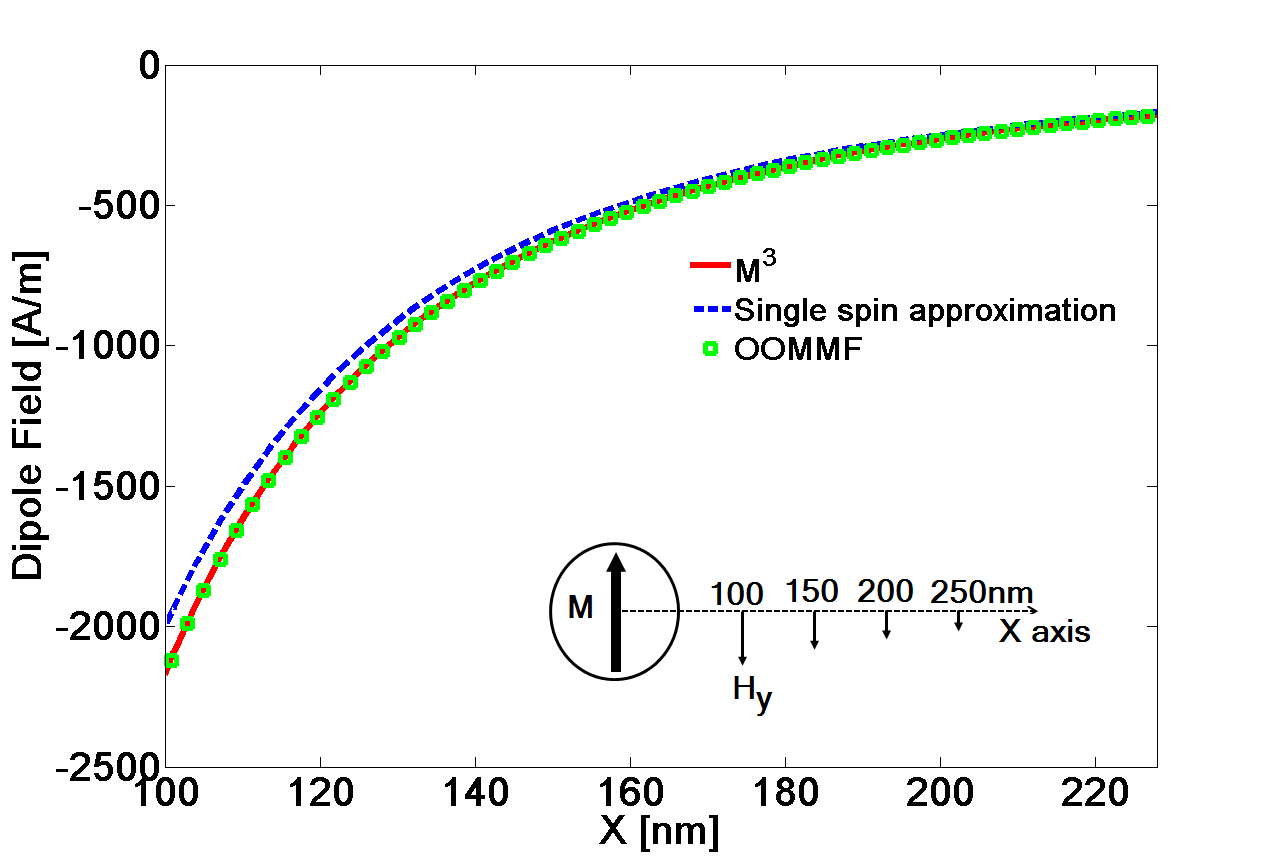,width=3.5in,height=2.5in}}
\caption{Dipole field for a single NM using OOMMF, M$^3$ and single 
domain approximation. Dipole field ($H_y$) is plotted along X axis \textcolor{black}
{which is the hard axis (or minor axis)} 
of the NM.} \label{band_structure}
\end{figure}

As information transfer from one NM to another takes place because of dipole 
coupling in NML, it is important to compare the dipole field of a multi domain system with 
single spin dipole approximation we use in our study where the magnet is assumed to be a 
single point (Eq. 6). The micromagnetic calculation is started with uniform magnetization 
across the NM and then relaxed it to its equilibrium configuration. Simulation is terminated 
when the residual torque satisfies $\vert \vec{m}\times\vec{H}\vert<10^{-4}$. Fig. A1 
shows the dipole field along the minor axis of the NM ($H_y$). At $d=150 {nm}$, the field 
from the OOMMF simulation is about $4.6\%$ larger than the single spin dipole approximation.
\textcolor{black}{At this distance, the gradient of the field is also small, so inhomegeneity 
of the dipole field over the second magnet's surface
is not significant.}

\begin{figure}[t]
\centerline{\epsfig{figure=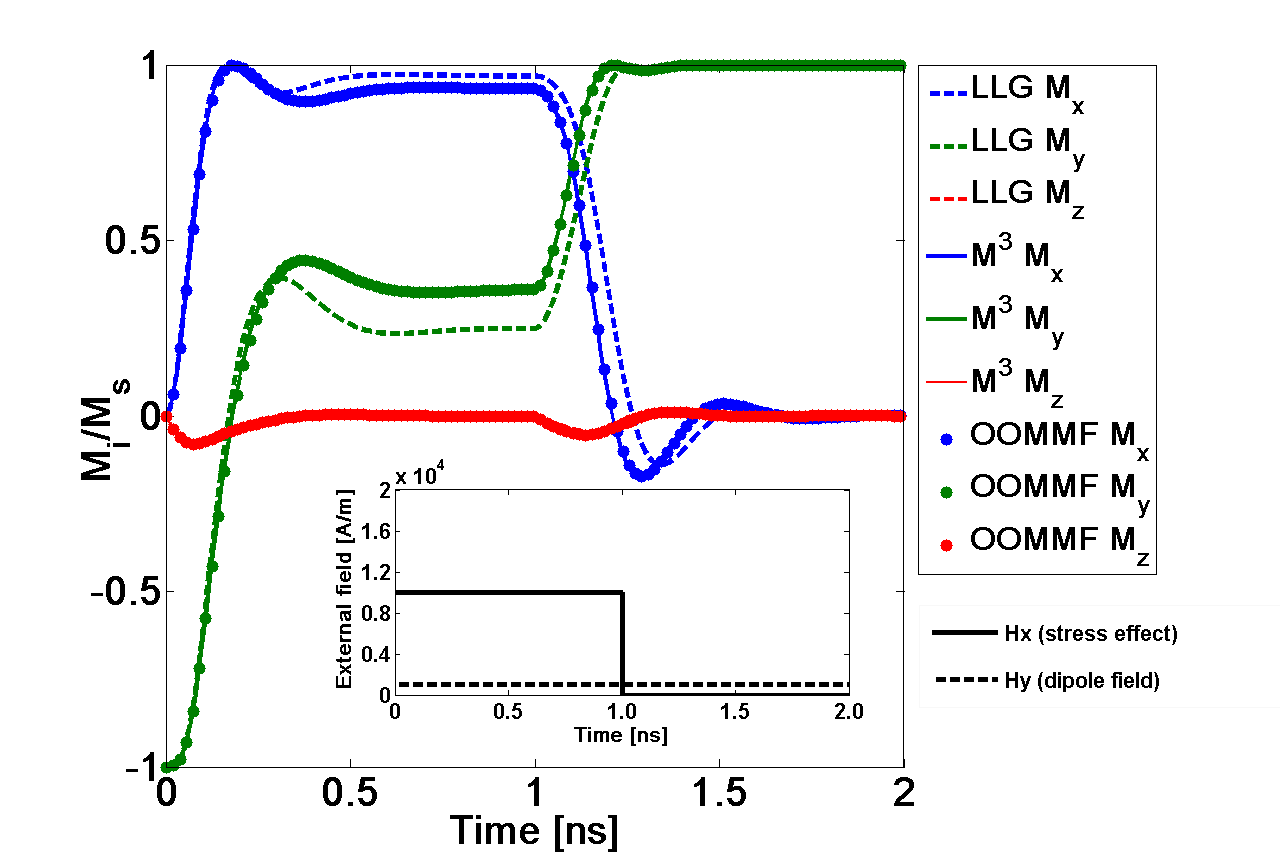,width=4in,height=2.5in}}
\caption{Magnetic simulation from OOMMF and M$^3$ and solution to the LLG 
equation for single domain magnet.} \label{band_structure}
\end{figure}

Since OOMMF/M$^3$ cannot incorporate the effect of stress \textcolor{black}{easily}, a 
magnetic field 
is used to induce switching and the results are compared with macrospin calculation. 
A steady field of 1042$Am^{-1}$ is applied along $-Y$ direction to simulate the dipole field 
from the adjacent magnet (calculated from a magnet with the same dimension at 120 nm distance). 
A sufficiently large magnetic field of 10 $KA m^{-1}$ is then applied in the $X$ direction for 
1 ns to simulate the stress. The magnetization of the adjacent magnet is assumed to be not affected 
by the switching of the magnet under study so its dipole field stays unchanged throughout the switching. 
Fig A1 shows the profile of external field and shows the comparison between micromagnetic simulation
(OOMMF and M$^3$) and single domain macrospin LLG simulation. The overall switching behavior is similar 
between the micromagnetic and LLG simulations. The slope of the switching curves  are in close agreement 
which implies same switching speed. For all three simulations, the magnetization aligns with the Y axis 
after about 1.2ns. The discrepancy in the intermediate stage might come from some degree of non-coherency 
captured by micromagnetic simulation or a small discrepancy in the dipole field as indicated in the previous 
paragraph.

In conclusion, we have shown that single domain macrospin assumption in our study for 
NML is a reasonable approximation.

\bibliographystyle{IEEEtran}
\bibliography{multiferroic}

\begin{thebibliography}{10}
\providecommand{\url}[1]{#1}
\csname url@samestyle\endcsname
\providecommand{\newblock}{\relax}
\providecommand{\bibinfo}[2]{#2}
\providecommand{\BIBentrySTDinterwordspacing}{\spaceskip=0pt\relax}
\providecommand{\BIBentryALTinterwordstretchfactor}{4}
\providecommand{\BIBentryALTinterwordspacing}{\spaceskip=\fontdimen2\font plus
\BIBentryALTinterwordstretchfactor\fontdimen3\font minus
  \fontdimen4\font\relax}
\providecommand{\BIBforeignlanguage}[2]{{%
\expandafter\ifx\csname l@#1\endcsname\relax
\typeout{** WARNING: IEEEtran.bst: No hyphenation pattern has been}%
\typeout{** loaded for the language `#1'. Using the pattern for}%
\typeout{** the default language instead.}%
\else
\language=\csname l@#1\endcsname
\fi
#2}}
\providecommand{\BIBdecl}{\relax}
\BIBdecl

\bibitem{bandy_review}
S.~Bandyopadhyay and M.~Cahay, ``Electron spin for classical information
  processing: a brief survey of spin based logic devices, gates and circuits,''
  \emph{Nanotechnology}, vol.~20, p. 412001, 2009.

\bibitem{Imre13012006}
A.~Imre, G.~Csaba, L.~Ji, A.~Orlov, G.~H. Bernstein, and W.~Porod, ``Majority
  logic gate for magnetic quantum-dot cellular automata,'' \emph{Science}, vol.
  311, no. 5758, pp. 205--208, 2006.

\bibitem{Cowburn25022000}
R.~P. Cowburn and M.~E. Welland, ``Room temperature magnetic quantum cellular
  automata,'' \emph{Science}, vol. 287, no. 5457, pp. 1466--1468, 2000.

\bibitem{roy:063108}
K.~Roy, S.~Bandyopadhyay, and J.~Atulasimha, ``Hybrid spintronics and
  straintronics: A magnetic technology for ultra low energy computing and
  signal processing,'' \emph{Applied Physics Letters}, vol.~99, no.~6, p.
  063108, 2011.

\bibitem{cmos}
``{ITRS} {R}eport {CORE9GPLL}\_{HCMOS9}\_{TEC}\_4.0 databook, {STM}icro,''
  2003.

\bibitem{roy:JAP}
K.~Roy, S.~Bandyopadhyay, and J.~Atulasimha, ``Energy dissipation and switching
  delay in stress-induced switching of multiferroic nanomagnets in the presence
  of thermal fluctuations,'' \emph{Journal of Applied Physics}, vol.~12, p.
  023914, 2012.

\bibitem{spedalieri}
F.~M. Spedalieri, A.~P. Jacob, D.~E. Nikonov, and V.~P. Roychowdhury,
  ``Performance of magnetic quantum cellular automata and limitations due to
  thermal noise,'' \emph{IEEE Transactions on Nanotechnology}, vol.~10, p. 537,
  2011.

\bibitem{6632926}
M.~Fashami, K.~Munira, S.~Bandyopadhyay, A.~Ghosh, and J.~Atulasimha,
  ``Switching of dipole coupled multiferroic nanomagnets in the presence of
  thermal noise: Reliability of nanomagnetic logic,'' \emph{Nanotechnology,
  IEEE Transactions on}, vol.~12, no.~6, pp. 1206--1212, Nov 2013.

\bibitem{Fashami2013}
\BIBentryALTinterwordspacing
M.~S. Fashami, J.~Atulasimha, and S.~Bandyopadhyay, ``Energy dissipation and
  error probability in fault-tolerant binary switching,'' \emph{Sci. Rep.},
  vol.~3, pp.~--, Nov. 2013. [Online]. Available:
  \url{http://dx.doi.org/10.1038/srep03204}
\BIBentrySTDinterwordspacing

\bibitem{roy:recent}
K.~Roy, ``Critical analysis and remedy of switching failures in straintronic
  logic using bennett clocking in the presence of thermal fluctuations,''
  \emph{Applied Physics Letters}, vol. 104, p. 013103, 2014.

\bibitem{PhysRevB.83.224412}
K.~Roy, S.~Bandyopadhyay, and J.~Atulasimha, ``Switching dynamics of a
  magnetostrictive single-domain nanomagnet subjected to stress,'' \emph{Phys.
  Rev. B}, vol.~83, p. 224412, Jun 2011.

\bibitem{roy:unpublished}
K.~Roy, J.~Atulasimha, and S.~Bandyopadhyay, ``Binary switching in a symmetric
  potential landscape,'' \emph{SCIENTIFIC REPORTS}, no. 3:3038, 2013.

\bibitem{bennett}
C.~H. Bennett, ``The thermodynamics of computation - a review,''
  \emph{International Journal of Theoretical Physics}, vol.~21, pp. 905--940,
  1982.

\bibitem{atulasimha:173105}
J.~Atulasimha and S.~Bandyopadhyay, ``Bennett clocking of nanomagnetic logic
  using multiferroic single-domain nanomagnets,'' \emph{Applied Physics
  Letters}, vol.~97, no.~17, p. 173105, 2010.

\bibitem{noel2011}
N.~D'Souza, J.~Atulasimha, and S.~Bandyopadhyay, ``Energy-efficient bennett
  clocking scheme for four-state multiferroic logic,'' \emph{IEEE Transactions
  on Nanotechnology}, vol.~11, pp. 418--425, 2012.

\bibitem{0957-4484-22-15-155201}
M.~S. Fashami, K.~Roy, J.~Atulasimha, and S.~Bandyopadhyay, ``Magnetization
  dynamics, {B}ennett clocking and associated energy dissipation in
  multiferroic logic,'' \emph{Nanotechnology}, vol.~22, no.~15, p. 155201,
  2011.

\bibitem{300668}
P.~K. Larsen, G.~L.~M. Kampschoer, M.~Van Der~Mark, and M.~Klee, ``Ultrafast
  polarization switching of lead zirconate titanate thin films,'' in
  \emph{Applications of Ferroelectrics, 1992. ISAF '92., Proceedings of the
  Eighth IEEE International Symposium on}, Aug 1992, pp. 217--224.

\bibitem{PhysRevLett.83.1042}
R.~P. Cowburn, D.~K. Koltsov, A.~O. Adeyeye, M.~E. Welland, and D.~M. Tricker,
  ``Single-domain circular nanomagnets,'' \emph{Phys. Rev. Lett.}, vol.~83, pp.
  1042--1045, Aug 1999.

\bibitem{dynamic}
G.~Bertotti, C.~Serpico, and I.~Mayergoyz, \emph{Nonlinear Magnetization
  Dynamics in Nanosystems}, Elsevier, Ed.\hskip 1em plus 0.5em minus
  0.4em\relax Elsevier Science, 2008.

\bibitem{Chikazumi}
S.~Chikazumi, \emph{Physics of Magnetism}, Wiley, Ed.\hskip 1em plus 0.5em
  minus 0.4em\relax Wiley, 1694.

\bibitem{0957-4484-23-10-105201}
M.~S. Fashami, J.~Atulasimha, and S.~Bandyopadhyay, ``Magnetization dynamics,
  throughput and energy dissipation in a universal multiferroic nanomagnetic
  logic gate with fan-in and fan-out,'' \emph{Nanotechnology}, vol.~23, no.~10,
  p. 105201, 2012.

\bibitem{1059598}
R.~Abbundi and A.~Clark, ``Anomalous thermal expansion and magnetostriction of
  single crystal {T}b.27{D}y.73{F}e2,'' \emph{IEEE Transactions on Magnetics},
  vol.~13, no.~5, pp. 1519 -- 1520, sep 1977.

\bibitem{PSSA:PSSA195}
K.~Ried, M.~Schnell, F.~Schatz, M.~Hirscher, B.~Ludescher, W.~Sigle, and
  H.~Kronmüller, ``Crystallization behaviour and magnetic properties of
  magnetostrictive {T}b{D}y{F}e films,'' \emph{physica status solidi (a)}, vol.
  167, no.~1, pp. 195--208, 1998.

\bibitem{0022-3727-41-16-164016}
J.~Walowski, M.~D. Kaufmann, B.~Lenk, C.~Hamann, J.~McCord, and M.~Manzenberg,
  ``Intrinsic and non-local gilbert damping in polycrystalline nickel studied
  by {T}i : sapphire laser fs spectroscopy,'' \emph{Journal of Physics D:
  Applied Physics}, vol.~41, no.~16, p. 164016, 2008.

\bibitem{OOMMF}
M.~J. Donahue and D.~G. Porter, ``{OOMMF} {U}ser's {G}uide,'' US Department of
  Commerce, Technology Administration, National Institute of Standards and
  Technology, 1999.

\bibitem{M3}
\BIBentryALTinterwordspacing
C.~Mewes and T.~Mewes. Matlab {B}ased {M}icromagnetics {C}ode {M}$^3$.
  [Online]. Available: \url{http://bama.ua.edu/~tmewes/Mcube/Mcube.shtml}
\BIBentrySTDinterwordspacing

\bibitem{sun2}
J.~Sun, ``Spin angular momentum transfer in current-perpendicular nanomagnetic
  junctions,'' \emph{IBM J. RES. \& DEV.}, vol.~50, no.~1, pp. 81--100, 2006.

\bibitem{brown}
G.~Brown, M.~A. Novotny, and P.~K. Rikvold, ``Langevin simulation of thermally
  activated magnetization reversal in nanoscale pillars,'' \emph{Physical
  Review B}, vol.~64, p. 134422, 2001.

\bibitem{5018466}
G.~Dewar, ``Effect of the large magnetostriction of terfenol-d on microwave
  transmission,'' \emph{Journal of Applied Physics}, vol.~81, no.~8, pp.
  5713--5715, Apr 1997.

\end{thebibliography}


\end{document}